\documentclass[journal]{IEEEtran}

\usepackage{graphicx}
\usepackage{url}
\usepackage[english]{babel}
\usepackage{blindtext}
\usepackage{multirow}
\usepackage[table]{xcolor}
\usepackage{footnote}
\usepackage{booktabs}
\usepackage{tabularx}
\usepackage{enumitem} 
\usepackage{siunitx} 

\begin{document}
\title{Aether: Network Validation Using Agentic AI and Digital Twin}


{
 }

\author{Jordan Auge, Sam Betts, Giovanna Carofiglio, Giulio Grassi, Martin Gysi, John Kenneth d'Souza
\thanks{J. Auge, S. Betts, G. Carofiglio, and G. Grassi are with Cisco Systems, Paris, France and London, UK.}
\thanks{M. Gysi and J. K. d'Souza are with Swisscom, Zurich, Switzerland.}
}


\maketitle

\begin{abstract}
    Network change validation remains a critical yet predominantly manual, time-consuming, and error-prone process in modern network operations. While formal network verification has made substantial progress in proving correctness properties, it is typically applied in offline, pre-deployment settings and faces challenges in accommodating continuous changes and validating live production behavior. Current operational approaches typically involve scattered testing tools, resulting in partial coverage and errors that surface only after deployment. 
    
    In this paper, we present \textit{Aether}, a novel approach that integrates Generative Agentic AI with a multi-functional Network Digital Twin to automate and streamline network change validation workflows.
    It features an agentic architecture with five specialized Network Operations AI agents that collaboratively handle the change validation lifecycle from intent analysis to network verification and testing. 

    Aether agents use a unified Network Digital Twin integrating modeling, simulation, and emulation to maintain a consistent, up-to-date network view for verification and testing. By orchestrating agent collaboration atop this digital twin, Aether enables automated, rapid network change validation while reducing manual effort, minimizing errors, and improving operational agility and cost-effectiveness.
   We evaluate Aether over synthetic network change scenarios covering main classes of network changes and on past incidents from a major ISP operational network, demonstrating promising results in error detection (100\%), diagnostic coverage (92-96\%), and speed (6-7 minutes) over traditional methods.
\end{abstract}

\section{Introduction}\label{sec:intro}

As networks grow in scale, complexity, and criticality, the ability to validate changes 
introduced by configuration updates and architectural modifications efficiently and reliably has become 
paramount for network operators in enterprise and service provider environments alike. 
Despite substantial research and technological advancements in the field, network change validation remains a largely manual, fragmented, and error-prone process, making it a contributor to network failures and the
nearly $400B$ of annual cost related to unplanned downtime~\cite{Oxford-Economics}.
Operational workflows typically rely on a patchwork of verification tools, simulation platforms, and ad-hoc testing solutions, each with limited scope and interoperability. This fragmented landscape not only increases the operational burden on skilled professionals but also leads to gaps in testing coverage, with errors frequently surfacing only after changes have been deployed into production networks.
The industry's growing adoption of automation and Dev-Ops-inspired methodologies (commonly referred to as NetDevOps) has alleviated some operational pain points by promoting repeatable workflows and incremental automation. However, 
the problem of providing comprehensive, end-to-end validation remains.
In parallel, the emergence of generative AI and agentic AI systems introduces new opportunities for automating and enhancing network operations. When combined with advances in network digital twin technology to provide dynamic, high-fidelity virtual replicas of production networks, there is significant potential to transform the way network change validation is performed.

In this paper, we introduce \textit{Aether}, a novel solution at the intersection of generative agentic AI, NetDevOps practices and network digital twins that combines the strengths of Agentic AI and multi-functional Network Digital Twins to deliver a unified, NetDevOps-automated and scalable network change validation solution integrated in existing processes.
Aether design is extensible both in terms of AI agents and of validation capabilities: an initial implementation has focused on verification and simulation functions and has been evaluated through practical case studies, including network change scenarios reproducing past incidents within a major ISP operational network. 
Aether benefits are quantified against the baseline by means of human and generative AI domain expertise by defining metrics that quantify issue detection accuracy, correctness, robustness and cost for analysis and testing.
Our analysis shows that Aether delivers promising results in automation, test coverage ( 92-96\%), error detection (100\%), and operational efficiency compared to existing practices, hence encouraging future development and extension of the approach.
This work does not raise any ethical issues.
The paper is organized as follows. Sec.~\ref{sec:rel_work} surveys network verification and early AI applications. Sec.~\ref{sec:prob_statement} presents the problem and baseline workflow. Sec.~\ref{sec:aether-architecture} details Aether's architecture (agents and NDT), with implementation in Sec.~\ref{sec:implem}. Secs.~\ref{sec:eval_methodo}-\ref{sec:evaluation} present methodology and results, including real-world ISP case studies (Sec.~\ref{sec:swisscom}). Finally, Secs.~\ref{sec:discussion}-\ref{sec:concl} summarize lessons learned and future work.


\section{Related work}
\label{sec:rel_work}

Network change validation has evolved through automation, preventative verification, and digital twins. This section surveys relevant research and tools, reviews early generative AI work, and motivates our approach.

\subsection{NetDevOps Automation}

Intent-driven NetDevOps automation is transforming network operations. Best practices now emphasize scripted automation, CI/CD pipelines, and Network Sources of Truth (NSOT) for managing configuration changes at scale. Tools like Ansible~\cite{Ansible}, Puppet~\cite{Puppet}, and configuration management databases enable configuration velocity and consistency. However, important challenges remain in both adoption and capabilities.

\textbf{Lack of semantic verification} - These systems validate configuration syntax and structural consistency, but fail to verify semantic effects on network state (operational consistency, i.e. whether the change breaks a critical policy).
    
\textbf{Limited testing capabilities} - Most automation solutions perform only generic tests that do not validate specific change intents. Meaningful validation often requires additional empirical testing through manually developed test plans, later integrated into test suites. This manual, error-prone, and reactive process scales poorly.
    
\textbf{Insufficient root cause analysis} - When tests fail, they signal problems without providing root cause analysis or guaranteeing that remediations address all side effects, perpetuating reactive troubleshooting.
To overcome such limitations, the research community has pioneered a more proactive and formal approach, often referred to as declarative network validation.
Here, we focus on previous work relevant to network change validation without attempting an exhaustive survey.

\subsection{Formal network verification}

Header Space Analysis (HSA) \cite{HSA} pioneered rigorous configuration analysis, aiming to mathematically prove network behavior properties rather than rely on empirical testing. Research from cloud operators including Microsoft Azure and Alibaba \cite{Azure1, Azure2, Azure3, netcov_sdn_coverage, Alibaba1, Alibaba2, Alibaba3, Alibaba4, rela} has been instrumental in translating fundamental research into solutions for real-world operational challenges in formal network verification (FNV). Early FNV efforts adopted abstract models independent of vendor-specific configurations. Declarative languages like Datalog \cite{Datalog} evolved to represent complex device packet-processing pipelines with sufficient fidelity for modeling vendor configurations. 

NetCov~\cite{netcov_sdn_coverage} introduced systematic metrics for quantifying test coverage in network verification, though technical coverage does not necessarily correlate with operational or business-critical concerns. Recent work by Krentsel et al.~\cite{hotnets25} on accessible model-free verification explores approaches to make formal verification more practical for operators.

Relational Verification \cite{rela} extended this work to cover change verification, comparing pre-change and post-change network snapshots to ensure modifications introduce only expected, localized effects—the approach we adopt in this paper. Relational NetKAT~\cite{xu2026network} recently introduced a compositional language to formally specify the mapping between pre- and post-change states, allowing operators to verify that end-to-end packet behavior evolves according to an intended transformation. Work to operationalise it remains open, and generative AI is a promising direction to bridge this formal modeling into CI/CD leveraging user intent.
Despite significant growth in open-source verification tools over the past decade, they remain largely underutilized in production environments, with notable exceptions like Batfish \cite{brown23lessons}.

\subsection{Network Digital Twins}

Network verification landscape spans multiple paradigms, each one with distinct characteristics and limitations, as summarized in Table~\ref{tab:verification_paradigms}.

\textbf{Model-based verification tools} represent one approach, with Batfish as a prominent HSA implementation. Batfish's strengths include what-if scenario analysis, formal model-based verification, and configuration ingestion across diverse vendor platforms. While Batfish excels at Layer 3 verifications—forwarding reachability, ACL (Access Control List) analysis, and model-based data-plane computation, it remains fundamentally limited to these domains. 
 
\textbf{Simulation-based platforms} bridge model-based and emulation approaches, offering dynamic analysis with statistical abstraction. Simulators like \cite{ns3}, \cite{omnet} provide varying fidelity levels, enabling realistic traffic patterns and protocol interactions while maintaining computational efficiency. Their main application is
performance analysis which requires scalable simulation of realistic traffic. Recently, tools like RouteNet \cite{rusek2020routenet} address scalability challenges by training ML models (Graph Neural Networks) for rapid performance prediction based on topology and traffic patterns. However, these specialized tools require complementary solutions like Batfish to compute the topology and forwarding plane states needed for what-if analysis.

\textbf{Emulation platforms} like~\cite{GNS3} or \cite{ContainerLab} pursue maximum fidelity through container-based virtualization, VMs, or native router images, exemplifying the core verification tradeoff: fidelity and scalability are inversely related, with hardware optimizations necessarily abstracted. While emulation captures vendor-specific behaviors and edge cases beyond model-based tools' reach, it requires substantially more compute resources and execution time. Physical network replicas offer ultimate fidelity for detecting otherwise-impossible errors, but their resource demands severely restrict verification scope and frequency.

Aether's design follows a modular compositional approach to network verification as recently proposed by the research community~\cite{NetKAT}.
Complex verification tasks are decomposed into smaller ones that can be independently analyzed and later combined to achieve end-to-end assurance. As an example, verifying an SRv6 overlay involves both examining the underlay’s forwarding behavior—using tools such as Batfish—and evaluating the overlay’s segment routing logic with dedicated SRv6-aware analyzers. Likewise, performance verification across multi-layer networks often requires integrating structural analysis from model-based tools with traffic-aware performance predictions provided by simulation tools like RouteNet.

\begin{table}[htbp]
\centering
\resizebox{0.85\linewidth}{!}{%
\begin{tabular}{@{}llll@{}}
\toprule
\small{\textbf{Type}} & \small{\textbf{Fidelity}} & \small{\textbf{Scale}} & \small{\textbf{Limitations}} \\
\midrule
\small{Model} & \small{Low-Med} & \small{High} & \small{L3 only, no traffic} \\
\small{Simulation} & \small{Medium} & \small{Medium} & \small{Needs topology} \\
\small{Emulation} & \small{High} & \small{Low} & \small{Resource intensive} \\
\small{Physical} & \small{Highest} & \small{Lowest} & \small{Limited scope} \\
\bottomrule
\end{tabular}%
}
\caption{Network Verification Paradigms}
\label{tab:verification_paradigms}
\end{table}
\textbf{Network Digital Twin Platforms}: To bridge the gap between different verification approaches, digital twin efforts (cfr.~\cite{ndt_survey} and references therein) have emerged to provide a more realistic, real-time mirror of the production network, serving as a sandbox for safe experimentation and test execution. The concept has gained significant traction lately within the Internet Engineering Task Force (IETF), where standardization efforts are underway to define architectural frameworks for Network Digital Twins (in IETF's Network Management Operations working group~\cite{simap})

Initiatives such as TwinEU~\cite{twineu} or TMForum Network Digital Twin group~\cite{TMForum} also explore standardized frameworks for defining digital maps and for interconnecting specialized digital twins across different network domains.


\subsection{Agentic Verification}

Applications of generative AI in networking have evolved from basic configuration generation~\cite{namrud2024kubeplaybook, sahoo2024ansible} to interactive assistants. Recent works like ``Ask Batfish''~\cite{mondal2023llms} demonstrate LLMs as efficient natural language interfaces for formal verification, effectively translating user queries into tool-specific commands. However, these systems function primarily as single-turn, reactive QA interfaces---requiring operators to know precisely \textit{what} to verify---or serve as static linters within CI/CD pipelines, lacking awareness of the specific change intent. 
Aether advances this state of the art by moving from reactive translation to proactive multi-agent orchestration. Unlike static CI/CD suites that run pre-defined checks, Aether's agents analyze natural language intent to dynamically generate and execute tailored verification plans, orchestrating heterogeneous tools to cover properties addressable only through tool composition (e.g., combining reachability with performance impact). This shift from "chatting with tools" to "autonomous validation workflows" enables handling the multi-step reasoning required for production change management.

\section{Problem Statement}\label{sec:prob_statement}

Drawing from related work analysis, we identify key limitations in current verification approaches and present Aether's design principles (\S\ref{sec:challenges}). We then establish the baseline network change validation process (\S\ref{sec:ncv_baseline}) that contextualizes Aether's operational setting.

\subsection{Motivation and Design Principles}\label{sec:challenges}

\textbf{Key Limitations.} As discussed in Sec.~\ref{sec:rel_work}, current verification approaches face three challenges: \textit{(i)} \textbf{Fragmentation}---verification tools present heterogeneous data models and incompatible interfaces across paradigms (model-based, simulation, emulation), forcing operators to manually coordinate multiple tools for comprehensive validation; \textit{(ii)} \textbf{Cognitive barriers}---operators face both the conceptual gap between natural language intent and abstract specifications (e.g., flow-set requirements~\cite{becket2020nv2}) and the operational complexity of using formal verification tools like Batfish, which require deep expertise to correctly specify queries and interpret results; and \textit{(iii)} \textbf{Agent operational requirements}---network configurations exceed LLM context windows, complex protocol interactions surpass reasoning capabilities, and agents require structured data access, specialized verification tools (for correctness and trust), scalable interfaces, common data models to present information uniformly to both humans and machines, and bidirectional natural language support to integrate into human workflows.

\textbf{Aether Approach.} Aether addresses these limitations through a neuro-symbolic architecture combining LLM-based agents with a unified Network Digital Twin (NDT) infrastructure. This separation enables flexible intent interpretation paired with correct, compositional verification:

\textit{(P1) Intent-aware compositional orchestration}: LLM agents (\S\ref{sec:aether-agents}) translate natural language change intent into verification workflows by compositionally orchestrating specialized tools---enabling differential verification tailored to change intent (e.g., verifying specific route absence rather than exhaustive reachability) that would be infeasible with static CI/CD test suites.

\textit{(P2) Unified infrastructure addressing agent operational requirements}: The NDT (\S\ref{sec:aether-ndt}) addresses agent scalability and reasoning limits through: (a) a temporal knowledge graph providing structured access to network data with query interfaces that fit LLM context windows; (b) specialized verification tools exposed via standardized APIs for correctness verification; (c) common data models (OpenConfig-based) presenting network state uniformly to agents and humans; and (d) natural language query support for graph traversal, enabling agents to gather targeted information without overwhelming their reasoning capabilities.

\textit{(P3) Workflow integration}: Support integration into existing NetDevOps practices through CI/CD hooks (GitHub Actions, network controller synchronization) and natural language interfaces aligned with established change management workflows (\S\ref{sec:ncv_baseline}).

Evaluation (\S\ref{sec:evaluation}) focuses on P1-P2, demonstrating intent-aware agent orchestration and compositional tool use. Analysis (\S\ref{sec:discussion}) examines agent-tool synergy and identifies that specialized knowledge injection is critical for handling protocol complexity. P3's CI/CD integration is illustrated through production use cases but not systematically evaluated.

\subsection{Network Change Validation process} \label{sec:ncv_baseline}
\begin{figure}
    \centering
    \includegraphics[width=0.6\linewidth]{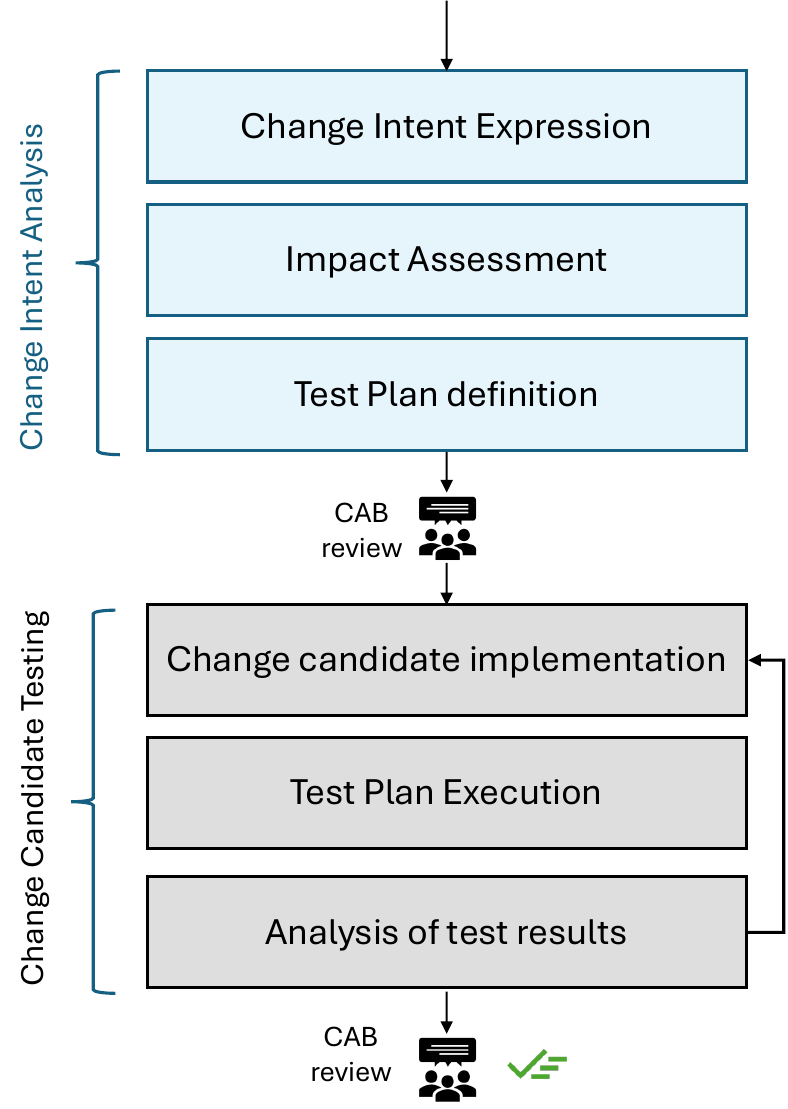}
    \caption{Network Change Validation process.}
    \label{fig:nvcp}
\end{figure}
Network change validation processes vary across domains and organizations but share common steps that we model as a baseline (Fig.\ref{fig:nvcp}). 
The process begins with documenting change intent in an IT Service Management (ITSM) tool, followed by impact assessment to evaluate potential effects on the network environment.
A test plan is then defined to validate all critical areas affected by the change. These artifacts—change intent, impact analysis, and test plan—undergo peer review by the Change Advisory Board (CAB) for initial approval.
Upon approval, the change candidate is developed and tested against the defined plan. Test results are analyzed and the candidate is iteratively refined as needed. Finally, a second CAB review examines the results and documentation before approving production deployment. This multi-stage process ensures thorough validation, risk mitigation, and service reliability.

\section{Aether}
\label{sec:aether-architecture}
Aether combines \textit{(i)} a unified Network Digital Twin (NDT)  with \textit{(ii)} a suite of specialized AI agents driving the network change validation process by interacting with the user and with the NDT. The NDT acts as the core representation and tool for network validation: it is composed of 
\begin{itemize}
    \item a \textit{Network Digital Map (NDM)}, i.e. a network representation in the form of a temporal graph allowing agents to access network status and knowledge in real-time and of
    \item a \textit{set of network verification and testing tools} operating on NDM time snapshots (production and candidate network states).
\end{itemize}
Aether generative AI Agents orchestrate validation workflows and automate decision-making throughout the change validation life-cycle.

\begin{figure}[h]
    \centering
    \includegraphics[width=0.9\linewidth]{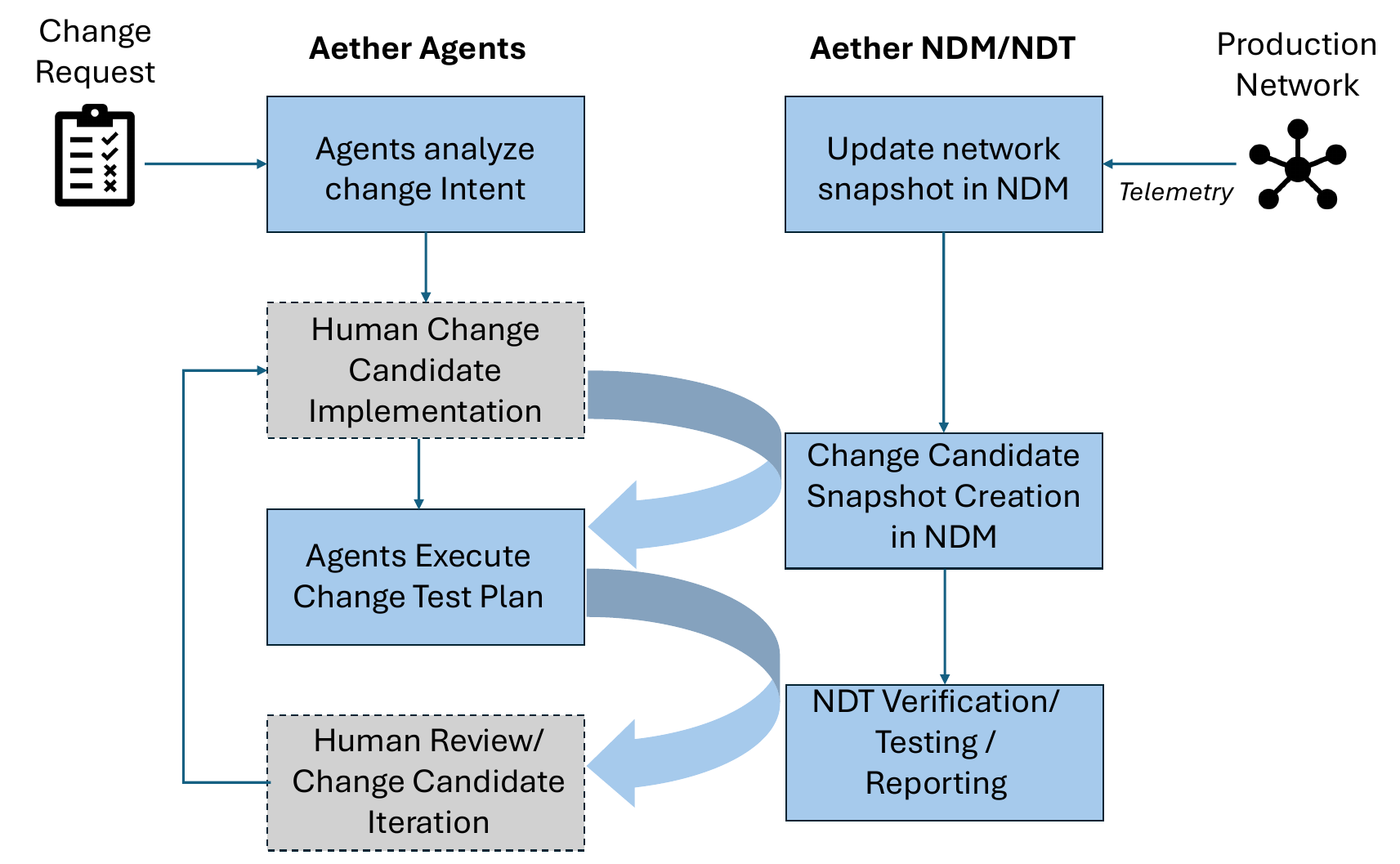}
    \caption{Aether workflow.}
    \label{fig:aether_overview}
\end{figure}
Aether workflow is described in Fig.\ref{fig:aether_overview}: network telemetry data from production network is periodically ingested into the NDM to maintain an accurate up-to-date representation of the current network state and configuration. When a change request is initiated, Aether agents collaborate to analyze the change intent, assess its potential impact and generate tailored test plans, based on the intent and the current state of the network. Once the change has been implemented (by a human or Aether-external automation), agents execute tests against the candidate change within the NDT environment. If test results indicate issues, the user may iterate on the change implementation and re-initiate the validation process. Once tests pass successfully, the change and validation results are presented for final approval; Aether is designed as a Human-in-the-Loop (HITL) system, ensuring that critical deployment decisions and risk acceptance remain under human operator authority.

The following subsections detail Aether's agent architecture (\S\ref{sec:aether-agents}) and Network Digital Twin (\S\ref{sec:aether-ndt}).

 \begin{figure}[h]
    \centering
    \includegraphics[width=\linewidth]{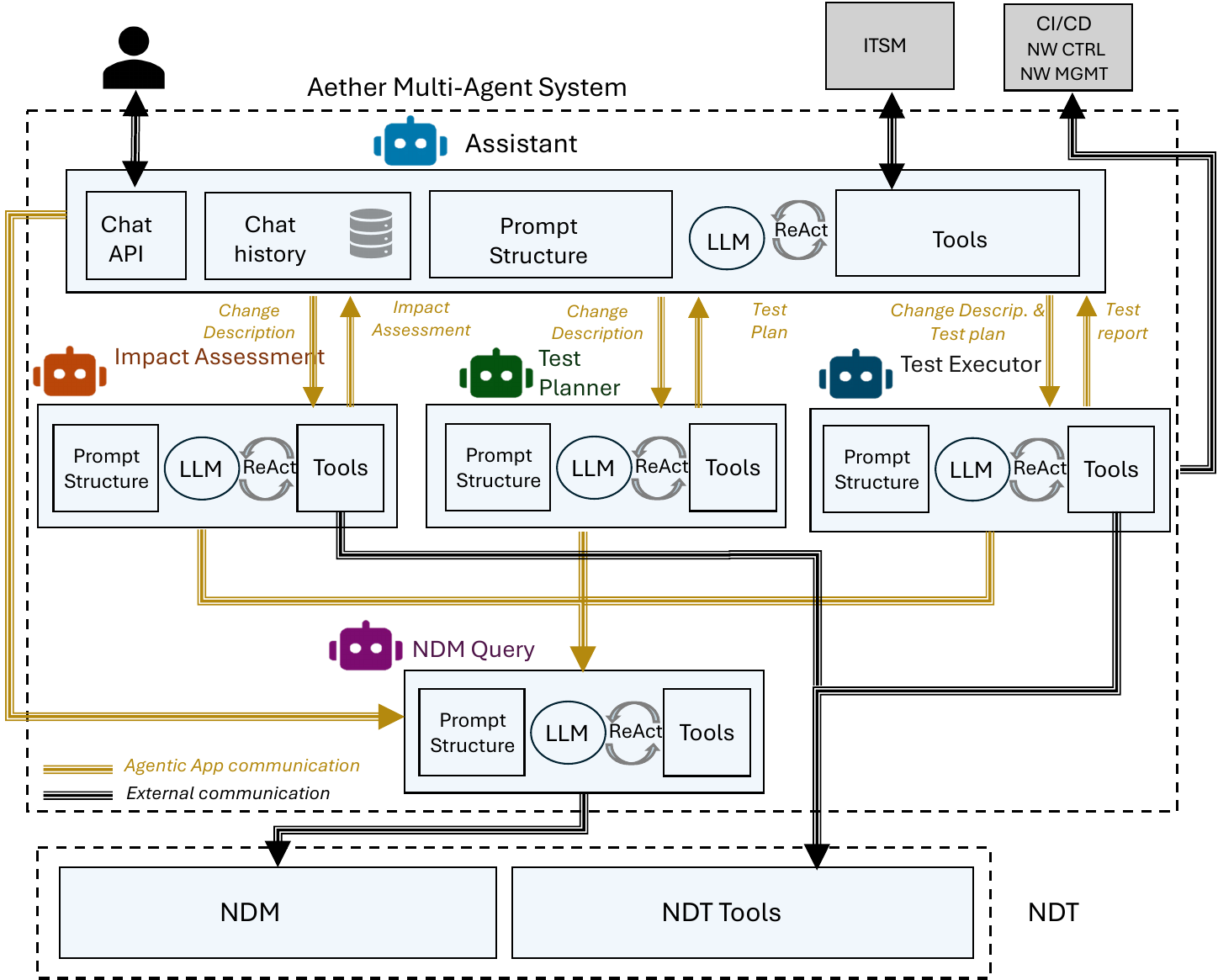}
    \caption{Agents logical architecture.}
    \label{fig:agents_logical_arch}
\end{figure} 
\subsection{Aether Agents}
\label{sec:aether-agents}

Aether employs a multi-agent architecture to automate network change validation, represented as external entities in Figure~\ref{fig:agents_logical_arch}. Agent roles map to the baseline change management workflow, covering: impact assessment, test planning, test execution and reporting. They are orchestrated by a conversational Assistant Agent that serves as the primary user interface. 
Such agents interact with Aether Network Digital Map (NDM) via an NDM Query agent allowing natural language queries to NDM Knowledge Graph data.

The multi-agent architecture provides the modularity and extensibility required by our design requirements. By decomposing validation into specialized agents, each focused on a distinct task, the system enables independent development, replacement, or augmentation of capabilities without disrupting the workflow. Task-specific agents maintain focused, domain-relevant context, enhancing reasoning accuracy and efficiency.

All agents support tool calling through the ReAct design pattern~\cite{react}, allowing  planning, iteration and self-correction based on tool answers. The Assistant (or orchestrating workflow) is responsible for dispatching to specialized agents, and provide them with all required input artifacts such as: (1) context (change intent, ticket id, pull-request id), and (2) relevant artifacts (impact assessment, test plan) which are all persisted in the ITSM; and (3) per-change relevant chat history (shared working memory, backed by a database).
Agents are specialized through their system prompt defining their roles and objectives, available tools, formatting guidelines for the LLM, and more importantly their set of skills and access to memories (see Table~\ref{tab:agent_specs}).
\begin{table}[htbp]
\centering
\caption{Aether Agents Architecture and Specifications}
\label{tab:agent_specs}
\begin{tabular}{@{}p{0.15\columnwidth}p{0.8\columnwidth}@{}}
\toprule
\textbf{Assistant} & \textbf{Pattern}: ReAct, \textbf{LLM}: GPT-4o \\
\midrule
\textbf{Objective} & Natural Language (NL) interface with users, services and agents; Orchestration. \\
\textbf{Tools} & ITSM; Config. change repository; All agents \\
\textbf{Skills} & Dispatch tasks; Output summary; External artifact retrieval \\
\textbf{Memories} & Chat history, Ticket \& Change information \\
\textbf{Data} & \textbf{In}:User Query; \textbf{Out}:Action,Reply; \textbf{Ctx}:Ticket ID 
\\
\toprule
\textbf{NDM Query} & \textbf{Pattern}: ReAct, \textbf{LLM}: GPT-4o \\
\midrule
\textbf{Objective} & Provide network state, NL into Graph Queries. \\
\textbf{Tools} & NDM (Schema \& AQL Query) \\
\textbf{Skills} & Network Schema Navigation (OpenConfig), Protocol Logic, Graph Query Generation/Optim., Data Access Resilience \\
\textbf{Memories} & NDM/NDT Production snapshot \\
\textbf{Data} & \textbf{In}: NL Question; \textbf{Out}: Query response (raw/NL); \textbf{Ctx}: Snapshot ID \\
\midrule
\textbf{Impact A.} & \textbf{Pattern}: ReAct, \textbf{LLM}: GPT-4o \\
\midrule
\textbf{Objective} & Analyze change impact (affected devices/ layers, downtime, performance). \\
\textbf{Tools} &  NDM (Schema \& AQL Query); NDT Impact Assessment helper \\
\textbf{Skills} & Change Intent Analysis, NDT layered structure, Network Schema Navigation \\
\textbf{Memories} & NDM/NDT production snapshot \\
\textbf{Data} & \textbf{In}: Change Description \textbf{Out}: Assessment Report; \textbf{Ctx}: Snapshot ID\\
\toprule
\textbf{Test Plan} & \textbf{Pattern}: ReAct; \textbf{LLM}: GPT-4o \\
\midrule
\textbf{Objective} & Generate test plans 
\\
\textbf{Tools} & NDM (Schema \& AQL Query)  \\
\textbf{Skills} & Change Intent Analysis, Protocol Logic, Automated Relevance Filtering, Operational Verif. \\
\textbf{Memories} & Change Intent/Impact assessment(ITSM); NDM Production snapshot \\
\textbf{Data} & \textbf{In}: Change Description; \textbf{Out}: Test Plan; \textbf{Ctx}: Production Snapshot ID\\
\toprule
\textbf{Test Exec.} & \textbf{Pattern}: ReAct, \textbf{LLM}: GPT-4o \\
\midrule
\textbf{Objective} & Execute test plans, Retrieve Change candidate, Act as Validation Gateway. \\
\textbf{Tools} & NDM (Schema \& AQL Query); NDT Tools \\
\textbf{Skills} & Operational Verif.; Protocol Logic; NDT Tool use\\
\textbf{Memories} & Change Intent (ITSM); Test plan (ITSM); Change implementation;  \\
\textbf{Data} & \textbf{In}: Change Description, Test Plan, Change Implementation Reference; \textbf{Out}: Test Results \& Report; \textbf{Ctx}: What-if Snapshot ID\\
\bottomrule
\end{tabular}
\end{table}

\subsection{Aether NDT}
\label{sec:aether-ndt}
Aether Network Digital Twin 
serves as the backbone of the Aether architecture, facilitating seamless interaction between Aether generative AI agents and the underlying network data and tools. 
Aether NDT maintains an up-to-date graph-based Network Digital Map (NDM). It
provides the necessary context and information for the verification tools, ensuring that they operate on accurate and up-to-date representations of the network state. Aether NDT fetches data from the NDM as needed and transforms it based on the requirements of the tools. It also provides computing capabilities to the NDM to enrich the internal network representation.

\begin{table*}[ht]
    \centering
    \begin{tabular}{@{}llll@{}}
        \toprule \small{
        \textbf{Capability}} &  \small{\textbf{Category}} &  \small{\textbf{Tool}} &  \small{\textbf{Description}} \\
        \midrule
       \small{MTU consistency} &  \small{KG-based} &  \small{NDM} &  \small{Check all network links have same MTU} \\
        \small{Reachability} &  \small{Model-based} &  \small{Batfish} &  \small{Check that a source can reach a destination IP address} \\
         \small{Differential Reachability} &  \small{Model-based} &  \small{Batfish} &  \small{Check if two snapshots have same reachability properties}\\
        \small{Loop detection} &  \small{Model-based} &  \small{Batfish} &  \small{Check for loops in the network topology} \\
         \small{Traceroute} &  \small{Model-based} &  \small{Batfish} &  \small{Simulate a traceroute from a source to a destination} \\
         \small{ACL filters} &  \small{Model-based} &  \small{Batfish} &  \small{Search ACL filters and outcomes} \\
         \small{ACL filters} &  \small{Model-based} &  \small{Batfish} &  \small{Compare two snapshots for ACL filters and outcomes} \\
         \small{SLA verification} & \small{Simulation-based} &  \small{NS-3 + custom code} &  \small{Check if SLA is respected given a traffic demand}\\
        \small{SLA verification} &  \small{GNN-based} &  \small{Routenet}~\cite{rusek2020routenet} &  \small{Check if SLA is respected given a traffic demand}\\
         \small{Configuration validation} &  \small{Pattern-based} &  \small{Diffy}~\cite{kakarla2024diffy} &  \small{Check if configuration files exhibit anomalies} \\ 
    \bottomrule
    \end{tabular}
    \caption{NDT Verification Capabilities and Tools 
    }\vspace{-2mm}
    \label{tab:ndt_tools}
\end{table*}
\subsubsection{Network Digital Map}
\begin{figure}
    \centering
    \includegraphics[width=0.9\linewidth]{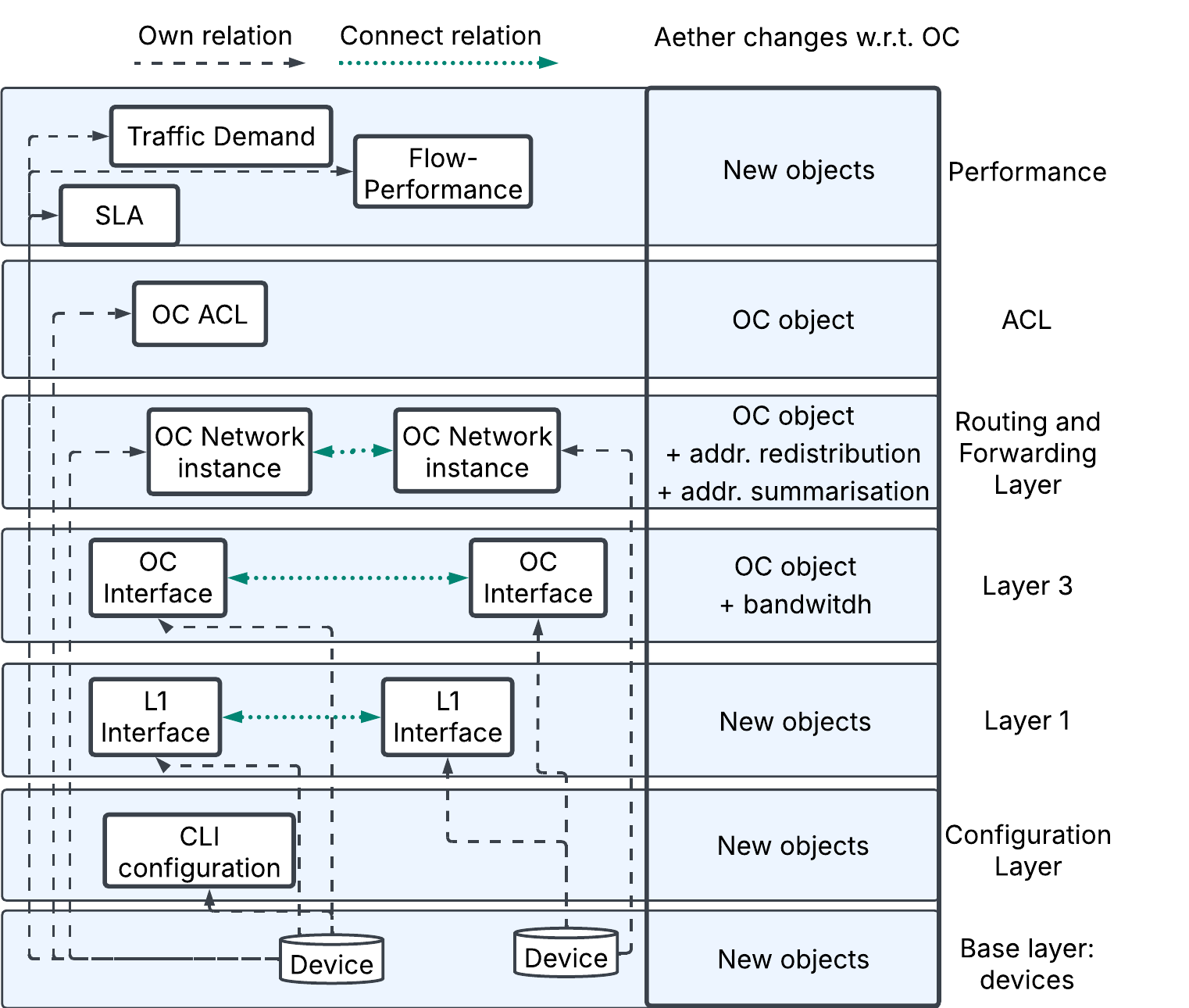}
    \caption{NDM Knowledge Graph structure. 
    }
    \label{fig:kg_details}
\end{figure}

The Network Digital Map (NDM) serves as a unified data model and network source of truth (NSOT) for Aether agents that aggregates and normalizes information from diverse network sources into a standardized schema, providing a comprehensive view of network state and configuration as a Knowledge Graph.
The data schema is based on OpenConfig~\cite{openconfig}, a widely adopted vendor-neutral standard for network configuration and management. We chose OpenConfig over alternatives like IETF YANG modules due to its stronger vendor adoption and focus on operational usability.
Aether NDM extends the OpenConfig schema to accommodate Aether NDT requirements, ensuring accurate capture of all relevant network attributes.
At the core of Aether NDM is a Knowledge Graph (KG) modeling relationships between network entities and attributes. The KG is structured into multiple layers, each corresponding to a specific network aspect. The base layer represents physical devices and basic attributes, while higher layers capture complex configurations such as routing policies, L3/L1 interfaces, ACLs~(respectively, the OpenConfig modules network-instance, interfaces and ACL are used), CLI configuration, and performance metrics. Each layer is self-contained and independent, enabling flexible integration and extension as requirements emerge. New OpenConfig modules can be added seamlessly.
Fig.~\ref{fig:kg_details} shows how NDM layers connect to form a KG. The base layer contains device nodes linked to other layers via \textsf{OWN} relations. At each layer, nodes describe specific device status and configuration aspects. Nodes across layers are linked via \textsf{CONNECT} edges, representing relationships between properties owned by different devices (e.g., network instances connected because routes exist between devices). 

\textbf{Data Ingestion in Aether NDM}
The ingestion in Aether NDM is architected as a modular pipeline with three main components: (1) a set of \emph{source adapters} that ingest input data from a specific source, process it, clean it, and normalize it to generate Aether-OpenConfig (Aether-OC) objects; (2) a \emph{KG builder} that, given a set of Aether-OC objects, constructs the Knowledge Graph by creating nodes and edges---this step may leverage NDT computing capabilities to enrich the graph with computed relationships or attributes; 
and (3) a \emph{graph database adapter} to store the KG (eg. ArangoDB~\cite{arangodb}). 
Aether NDM supports extraction of network data from a variety of sources, including direct device access via Netconf~\cite{enns2011rfc}, controller APIs (e.g., Cisco Network Services Orchestrator~\cite{cisco_nso_2024}), and simulation or analysis tools (e.g., Batfish). 
It supports ingestion from production networks as well as from change candidates: for the latter, it exploits the NDT computing capabilities to generate the required KG layers starting from the new device configurations i.e. it uses Batfish to normalise vendor-specific configurations and compute routing/forwarding information.

The types of data ingested encompass OpenConfig (OC)-like 
structured configuration and state, raw device configuration, CLI command outputs, and flow performance metrics.

\textbf{Building Aether NDM Knowledge Graph}
Constructing a full KG for every network snapshot may be resource-intensive and often unnecessary. To optimize, Aether allows on-demand ingestion: only essential layers are built initially, and original input data is stored separately. Additional layers are generated as needed for specific verifications or queries and merged to the existing KG. The \emph{database-adapter} merges new layers into the existing KG with per-layer granularity, avoiding duplication. 

\textbf{NDM schema API for agents}
The NDM Query Agent requires precise and unambiguous information about the structure and semantics of the Knowledge Graph (KG). Public OpenConfig documentation, while useful, is often too broad, may include unused modules, and does not reflect Aether-specific extensions.
To enable the agent to construct correct and efficient queries, the NDT
provides a dedicated
schema API that exposes the structure of the KG, as well as the schema of each type of nodes and edges, with examples, descriptions, and example-queries tailored to the KG-database. 
\subsubsection{Aether NDT Tools}
\label{sec:aether-ndt-tools}
Aether Network Digital Twin provides a unified platform integrating multiple network verification and testing tools for comprehensive automated validation. It serves as the backbone of the Aether architecture, facilitating seamless interaction between generative AI agents and underlying network data and tools.
Aether NDT currently supports model-based verification (Batfish~\cite{brown23lessons}) and simulation-based tools (Routenet~\cite{rusek2020routenet}), with emulation support planned. Each tool integrates through a standardized interface, allowing agents to invoke capabilities without understanding underlying complexities. The extensible design enables adding new tools while maintaining a consistent agent interface. Table~\ref{tab:ndt_tools} summarizes the main verification capabilities.
The NDT maintains close integration with the Network Digital Map (NDM), which provides necessary context ensuring tools operate on accurate, up-to-date network state. The NDT fetches and transforms NDM data based on tool requirements and provides computing capabilities to enrich the network representation. For traffic SLA verification, two options are available—simulation-based (Perf-simulator) and GNN-based~\cite{rusek2020routenet}—allowing selection based on scenario requirements (accuracy vs. scalability).

\textbf{Tool Exposure and Execution.}
Tools are exposed to agents via a Model Context Protocol (MCP) server, which provides input schemas and tool descriptions templated into the LLM context with tool-specific tokens. Tool descriptions are based on underlying tool documentation (e.g., Batfish) and modified for agent comprehension and NDT abstractions.
The test executor agent leverages NDT verification capabilities to implement test plans. It selects appropriate verifications based on requirements, uses the NDM Query Agent to gather necessary information (e.g., device IP addresses), executes verifications via the NDT, and retrieves results for analysis.
For Batfish-based verifications, the NDT translates agent input into Batfish-specific parameters and normalizes results. MTU consistency verification directly queries the NDM-KG for link MTU values. SLA verification involves multiple steps: simulating per-flow traffic matrices based on demand models (stored in NDM-KG), running the selected tool (NS-3 or Routenet) to evaluate flow performance (loss, delay), and comparing results against SLA requirements (also in NDM-KG). 

\begin{figure}
    \centering
    \includegraphics[width=1\linewidth]{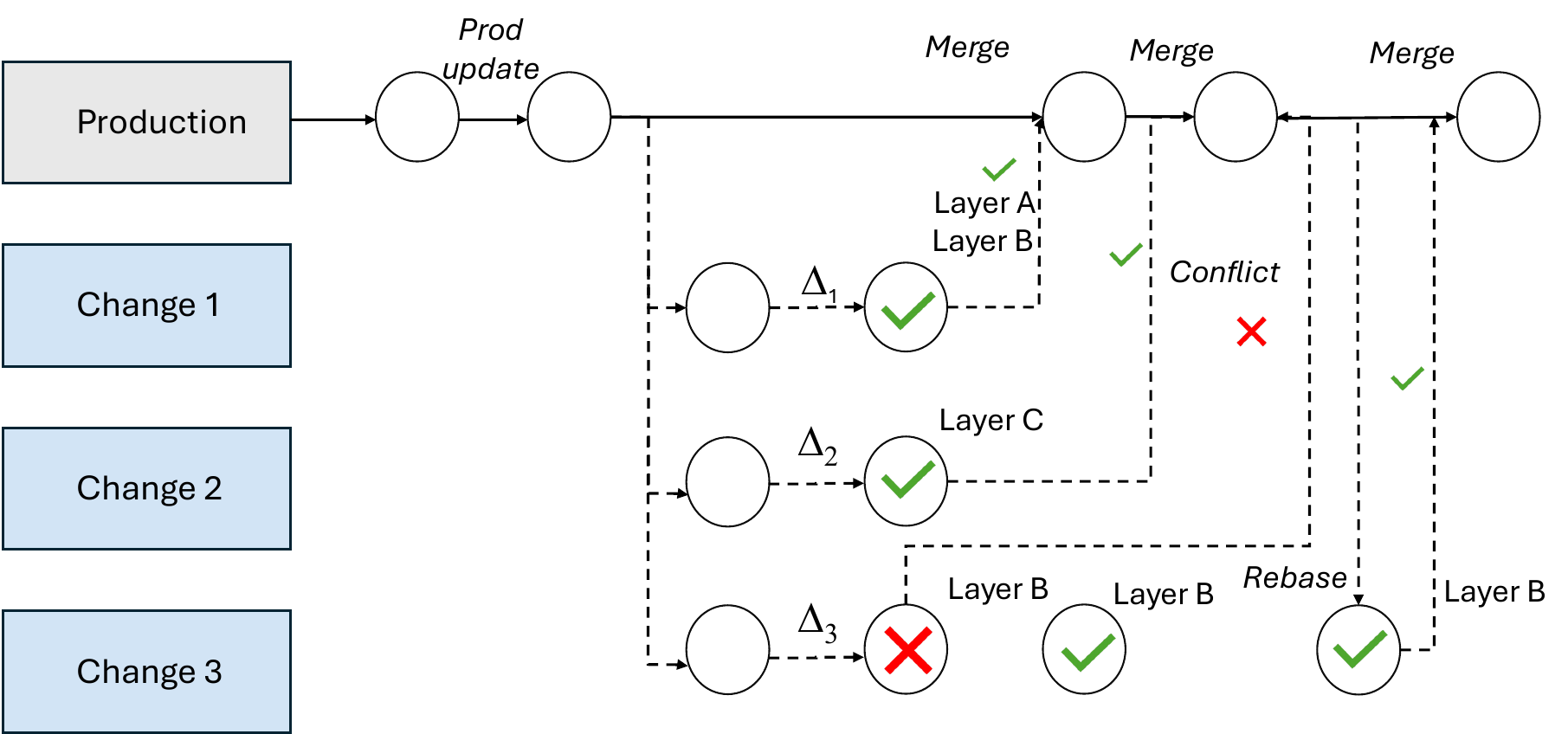}
    \caption{NDT workflow to manage snapshots changes.}
    \label{fig:git_workflow}
\end{figure}

\textbf{Git-like Workflow.}
The NDT uses snapshots as consistent network views at specific points in time, identified by UUIDs analogous to git commits. Production evolves along a "main" branch, with snapshots created for Pull Requests to validate changes before merging.
The layered KG structure enables efficient conflict resolution through per-layer evaluation using computed digests (hashes) per device and layer. Only affected layers require re-validation when conflicts occur, minimizing overhead. Layer dependencies (e.g., ACL changes affecting routing) ensure dependent layers are re-evaluated when parent layers change.
Figure~\ref{fig:git_workflow} illustrates this: Snapshot 1 applies change $\Delta$1 to layers A and B, passes verification, and merges. Snapshot 2 applies $\Delta$2 to independent layer C and merges successfully. Snapshot 3 applies $\Delta$3 to layer B, fails verification, is corrected to $\Delta$3' and passes, but requires rebase when production advances to Snapshot 4 with conflicting layer B changes.
The NDT exposes git-like APIs for fetching snapshots, forking, updating with change candidates, running verifications (including comparisons), and handling conflicts/rebases, enabling automated end-to-end change validation mirroring familiar git workflows.

\section{Aether Implementation}
\label{sec:implem}

\textbf{Agents} -- Aether leverages state-of-the-art agentic protocols defined by emerging bodies~\cite{AAIF, agntcy} to ensure interoperability and standard compliance. Agents are implemented in Python using the \textsf{A2A SDK}~\cite{a2a_spec_2025} and \textsf{LlamaIndex}~\cite{liu2023llamaindex}, utilizing the ReAct reasoning loop for tool orchestration. All agents use gpt-4o (Azure OpenAI); initial experiments with smaller models revealed limitations in tool utilization and schema adherence. 

Communication relies on the \textsf{A2A protocol} over the \textsf{SLIM} (Secure Low-latency Interactive Messaging)~\cite{SLIM} layer. The Assistant Agent orchestrates specialized agents while supporting direct peer-to-peer interactions (e.g., Test Planner querying NDM Query agent). NDT tools are exposed through an \textsf{MCP} (Model Context Protocol) server~\cite{mcp_spec_2025}, decoupling tool lifecycle from agent logic.

\textbf{NDT} -- Aether NDT is implemented in Go (REST APIs for agents, gRPC for NDM integration). The NDM uses Python-based source adapters and KG builder with ArangoDB for graph storage and operations, exposing gRPC APIs for ingestion and updates. Verification capabilities are exposed to agents via MCP server with tool descriptions and schemas.

\textbf{NDT Tools} -- We extended Batfish for IPv6 support, SRv6 locator origination, multiple ISIS processes per VRF (IOS-XR), intra-ISIS redistribution, and prefix summarization. 

For simulation-based performance analysis, we integrated two tools: a custom flow-level simulator built on NS-3, and RouteNet for ML-based performance prediction—both extracting topology and traffic demands from the NDM. Our simulation workflow reconstructs topologies and traffic from NDT layers, executes multiple runs, and aggregates metrics—establishing a reusable pattern that extends to emulation workflows with containerized network instances. 

Diffy~\cite{kakarla2024diffy} was adapted for differential anomaly detection across snapshots, applied to NDM-generated JSON configurations with anomaly-to-configuration mapping.

\textbf{Deployment and scalability} -- All components deploy via HELM/Docker in Kubernetes for lifecycle management and scaling. NDT introduces a thin dispatching layer with low overhead relative to verification resource consumption -- crucial for decoupling large configurations from agent context and uniformizing API queries. 

NDM scaling relies on ArangoDB for both ingestion and querying. For ingestion, leveraging existing network orchestrators (e.g., NSO~\cite{cisco_nso_2024} in our production use cases) enables efficient data collection and pre-normalization at scale, avoiding the need to poll individual devices directly.

\section{Evaluation methodology}
\label{sec:eval_methodo}

\begin{table*}[t]
\centering
\begin{tabular}{@{}lllll@{}}
\hline 
\textbf{ \small{Scenario}} &  \small{\textbf{Problem}} &  \small{\textbf{Layer}} &  \small{\textbf{Context}} &  \small{\textbf{Verification}} \\
\hline
 \small{S1. Router Maintenance} &  \small{Backup path error} &  \small{L3 Routing} &  \small{Maintenance} &  \small{Model + Differential} \\
 \small{S2. Firewall Update} &  \small{Policy violation} &  \small{L3/L4 Security} &  \small{Evolution} &  \small{Model + Differential} \\
 \small{S3. ACL Refactoring} &  \small{Equivalence failure} &  \small{L3/L4 Security} &  \small{Refactoring} &  \small{Model + Invariant} \\
 \small{S4. MTU Consistency} &  \small{Device mismatch} &  \small{L1 Physical} &  \small{Maintenance} &  \small{Cross-device} \\
 \small{S5. VLAN Migration} &  \small{Multi-device coordination} &  \small{L2/L3} &  \small{Migration} &  \small{Multi-device} \\
 \small{S6. ISP Link Migration} &  \small{SLA violation} &  \small{Cross-layer} &  \small{Migration} &  \small{Compositional} \\
 \small{S7. IS-IS Redistribution} &  \small{Control plane loop} &  \small{L3 Control} &  \small{Evolution} &  \small{Model + Convergence}\\
 \small{S8. Route Summarization} &  \small{Forwarding blackhole} &  \small{L3 Control} &  \small{Migration} &  \small{Differential} \\
\hline
\end{tabular}%
\caption{Network Change Validation Scenarios}
\vspace{-2mm}
\label{tab:scenarios}
\end{table*}
We evaluate Aether agentic network change validation system through two complementary approaches: (1) a benchmark suite of network change scenarios designed to cover common operational tasks and failure modes; (2) operational case studies based on real-world ISP production incidents. 

To the best of our knowledge, this paper is the first to attempt a systematic assessment of agentic AI contributions to network change validation and lack of publicly available datasets makes a comparison with NetCov\cite{netcov_sdn_coverage}/RELA\cite{rela} not possible.

To allow future comparison, we define an evaluation methodology which we will publicly release along with scenarios, datasets and ground truth built by human experts.

\subsection{Network Change scenarios}\label{sec:scenarios}
Our benchmark dataset includes $8$ network change scenarios (Table~\ref{tab:scenarios}) spanning a range of operational tasks (router maintenance, policy updates, topology migrations) and complex failure modes (protocol logic errors, cross-layer interactions, latent path defects). The scenarios  are selected based on two key criteria: \emph{operational relevance}—focusing on common and critical failure types observed in production networks; and \emph{diagnostic diversity}—ensuring coverage across different validation techniques and network layers.
\\\textbf{Operational Relevance}.Three types of failures are covered:

\textbf{(1) Policy misconfigurations (S1, S2, S3):} ACL and firewall errors are the largest category of connectivity failures in large-scale cloud providers and remain pervasive in enterprise networks due to frequent policy evolution~\cite{becket2020nv2}. We include scenarios testing unintended side-effects of rule additions (S1), collateral damage from policy updates (S2), and semantic equivalence violations during refactoring (S3).

\textbf{(2) IGP logic errors (S7, S8):} While BGP misconfigurations are well-studied causes of Internet outages~\cite{mahajan2002bgw}, errors within Interior Gateway Protocols represent a critical yet under-examined failure domain. We address this gap with scenarios for control-plane loops from IS-IS redistribution (S7) and forwarding blackholes from route summarization (S8)---subtle protocol-specific bugs that evade generic reachability tests.

\textbf{(3) Latent path defects (S1):} Production networks commonly contain dormant failures in backup paths that remain undetected until failover events~\cite{oppenheimer2003internet}. Our router maintenance scenario targets proactive validation of rarely-exercised paths before production activation.

\subsection{Evaluation Datasets}
We evaluate Aether agents first individually, then end to end in terms of agentic workflow. To this aim and in the absence of public datasets, Aether evaluation datasets are curated by a network expert with generative AI assistance: for each network scenario expert-generated input and ground truth output are defined. The test matrix consists of: (i) one input per scenario, (ii) three LLM-generated natural language input variations to test \textit{robustness} to natural language phrasing, (iii) one correct and one problematic change candidates, and  (iv) $10$ independent runs on all combinations to assess \textit{consistency}. 

For the NDM query agent specifically, the evaluation dataset includes $50$ questions LLM-generated from knowledge graph data related to the considered network scenarios covering all KG layers (Interfaces, IP settings, Routing/forwarding, ACL, device configuration). Each question is tested along with three prompt variations across $10$ independent runs.

\subsection{Evaluation Metrics}\label{sec:metrics}
Aether's agentic performance is quantified by the evaluation metrics in Table~\ref{tab:metrics}: 
(1) \textbf{Change candidate classification precision} or simply \textbf{Precision}, measuring the ability of the system to correctly validate a change, where $TP$ (True Positives) are faulty changes that the system blocks (at least one test fails) and $FP$ (False Positives) are valid changes that are blocked (at least one test fails).
 Higher precision indicates better diagnostic capability in identifying real issues while avoiding false alarms;
 Similarly, we define (2) \textbf{Main Error classification precision}, the metric that measures the ability of the system to detect the main problem in a bad change candidate: a change is considered correctly diagnosed only if the test that can catch the main problem fails for a faulty change ($TP$) and succeeds for a valid change ($TN$);
(3) \textbf{Error Detection}, measuring the ability of the system to detect broken changes, where $TN$ (True Negatives) are valid changes that go through (all tests pass); (4) \textbf{Time to Answer}, the average time taken by Aether to produce a response.
For the real-world scenarios, we further explore the intermediate test plan quality by defining $R_{GT}$ as the set of ground-truth validation requirements and $T_{gen}$ as the set of agent-generated tests. We measure:
(5) \textbf{Coverage} as the proportion of requirements satisfied by at least one test (denoted $R_{cov}$);
(6) \textbf{Efficiency} as the ratio of useful tests, defining $T_{relevant}$ for tests mapping to valid requirements (thus filtering out hallucinations); and 
(7) \textbf{Redundancy} quantifying how many generated tests target the same intent.

Finally, for the single-agent evaluation we define:
 (8) \textbf{Correctness}, assessed using GEval within the Deepeval framework~ \cite{deepeval}, which allows defining specific scoring criteria for the LLM judge: agents' output is compared against the ground truth and instructions are provided to the judge to focus on the facts and score based on the criteria reported in Figure~\ref{fig:single_agent_eval}; (9) \textbf{Robustness}, quantifying the agent's sensitivity to input variations, calculated based on $\sigma$, the standard deviation of correctness scores across test variations and $\mu$, the mean correctness score (i.e. a robustness score of 1.0 indicates perfect consistency); and (10) \textbf{Consistency}, measuring stability across repeated runs under identical conditions, also derived from correctness variance.

\begin{table}[h]
\centering
\renewcommand{\arraystretch}{1.3}
\begin{tabular}{@{}lp{2.9cm}p{3cm}@{}}
\hline
\textbf{Metric} & \textbf{Formula} & \textbf{Description} \\
\hline
Precision & $\frac{TP}{TP+FP}$ & Change candidate classification \\
Error Detection & $\frac{TP}{TP + FN}$ & Faulty candidates detection \\
Time to Answer & Average (sec) & Execution time \\

\hline
Coverage & $|R_{cov}| ~/~ |R_{GT}|$ & Expert requirements met \\
Efficiency & $|T_{relevant}| / |T_{gen}|$ & Ratio of useful tests \\
Redundancy & $1 - |R_{cov}|/ |T_{relevant}| $ & Duplication/overlap \\
\hline
Correctness & LLM score $\in [0, 1]$ & Alignment with ground truth \\
Robustness & $1 - \frac{\sigma}{\mu}$ (variations) & Stability across prompts \\
Consistency & $1 - \frac{\sigma}{\mu}$ (runs) & Stability across runs \\
\hline
\end{tabular}
\caption{Evaluation Metrics}
\label{tab:metrics}
\end{table}

\section{Evaluation}
\label{sec:evaluation}
We report now on the evaluation of Aether:
individual agents (Sec.~\ref{sec:single-agent-eval}), then end-to-end system (Sec.~\ref{sec:E2E}).
\subsection{Single Agent Assessment}
\label{sec:single-agent-eval}
\begin{figure}
    \centering
    \includegraphics[width=1\linewidth]{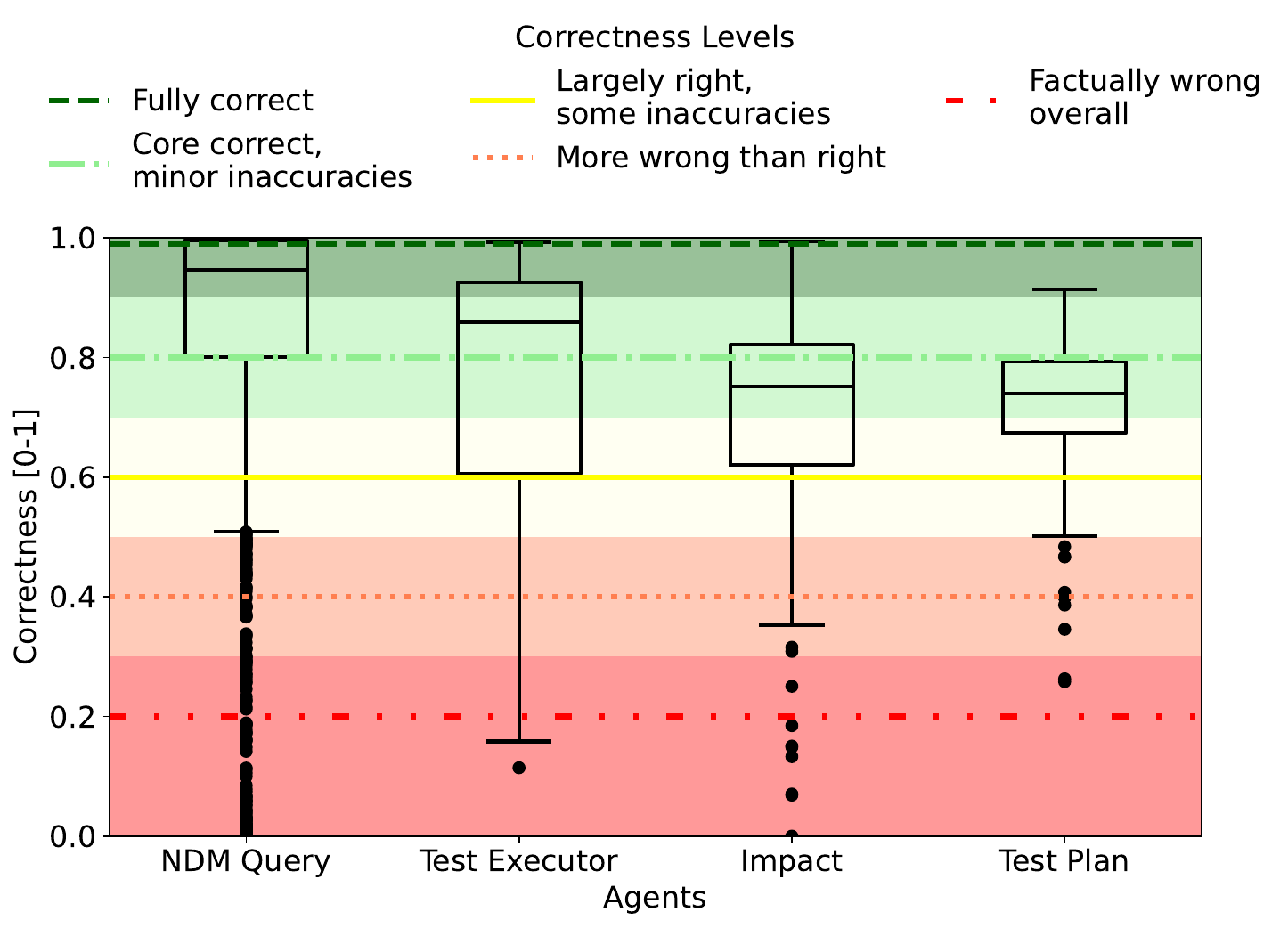}
    \caption{Single agent correctness.}
    \label{fig:single_agent_eval}
        \vspace{-2mm}
\end{figure}

Figure~\ref{fig:single_agent_eval} reports the correctness score averaged over each test across all scenarios, with thresholds defined in the LLM-as-a-Judge (LaaJ) prompt.
The average correctness score is always higher than the correctness threshold $0.7$, indicating Aether agents provide correct answers with only minor inaccuracies or missing non-critical information.
The \textbf{NDM Query agent} shows the highest average correctness with complete answers in most cases. However, a small subset of challenging queries causes the majority of failures, particularly those requiring deep OpenConfig structure knowledge and networking concepts (import/export routing policies, BGP reflection, static routes, ACL specifics). The data schema offers multiple expression paths for the same concept, and the agent often picks the option lacking KG data without exploring alternatives. Improvements can be obtained by enhancing query exploration, memory-based learning, or populating alternative paths at ingestion time. Other error sources include wrong attribute selection and logic errors.
The \textbf{Test Executor agent}'s main failure mode is misuse of NDT tool parameters, suggesting API simplification could help. Despite occasional mistakes, the agent performs correctly in most scenarios.
The \textbf{Impact Assessment agent} occasionally misses information due to task generality, though it provides good device and risk overviews. User interaction and ticket updates can clarify missing information.
The \textbf{Test Planner agent} shows satisfactory correctness, with core plans matching ground truth while occasionally missing tests for less common issues or including unnecessary tests.
The evaluation focuses on NetOps agents and does not consider in isolation the Assistant Agent, given that its primary function is to interface with the user and coordinate other agents. Its performance is assessed in the end-to-end scenarios.

Table~\ref{tab:single_agent_eval_overall} summarizes average Consistency, Robustness, and Time to Answer across scenarios. Overall values are high, indicating stable answers across runs and prompt variations. The Test Executor shows lowest consistency due to: (1) ground truth listing only one valid solution when multiple exist, and (2) NDT parameter misuse causing drastic score decreases when tests unexpectedly fail.
Average response times are acceptable at under two minutes for most agents. The Test Execution Agent takes longer due to actual test execution in the NDT.
\begin{table}[ht]
\centering
\begin{tabular}{ c c c c }
\hline
\textbf{Agent} & \textbf{Consistency} & \textbf{Robustness} & \textbf{Time to Answer} \\
\hline
NDM Query  & 83.3 & 82.2 & 12.6 \\
Test Executor & 73.8 & 92.2 & 89.0 \\
Impact & 78.0 & 93.7 & 46.7 \\
Test Plan & 83.7 & 95.1 & 46.8 \\
\hline
\end{tabular}
\caption{Single agent evaluation.}
\vspace{-5mm}
\label{tab:single_agent_eval_overall}
\end{table}

\subsection{End-to-end performance}
\label{sec:E2E}
In this section we evaluate the ability of Aether to properly validate a change candidate (Precision) and more specifically to block broken changes (Error Detection).
Table~\ref{tab:e2e_agent_eval_overall} reports the per-scenario Error Detection and Precision as well as the average results across all scenarios. Overall, Aether is able to block $94\%$ of the broken changes (Error Detection = 0.94) and to provide 0.64 precision in validating the outcome of the change candidate. 
When focusing on the detection of the main problem instead, the precision increases from 0.64 to 0.89 (Main Error Precision): the test plan is quite broad in covering different aspects of the change, leading to some false positives, especially when the tests require data not fully supported or available in the NDM. These are normally not related to the main problem introduced in the broken changes, hence the higher Main Error Precision.
Overall, the results show that Aether can accurately detect the most relevant and critical problems, enabling streamlined network change validation. The user remains in the loop and may validate Aether's actions by analyzing the detailed final report and updating the ITSM ticket to improve test plan and verifications.
Guardrails, agent monitoring as well as improving the robustness of the NDM Query Agent and extending the NDM data coverage are the next steps towards making Aether more autonomous and production-grade.
%
\begin{table}[ht]
\centering
\begin{tabular}{c c c c}
\hline
\textbf{Scenario} & \textbf{Error } & \textbf{Precision} & \textbf{Time to}\\
\textbf{n.}& \textbf{Detection} & \textbf{All - Main Error} & \textbf{Answer}\\
\hline
Scenario 1 & 0.6 & 0.67 - 0.78 & 245\\
Scenario 2 & 0.9 & 0.56 - 0.7 & 163 \\
Scenario 3 & 1.0 & 0.62 - 1 & 269\\
Scenario 4 & 1.0 & 0.62 - 0.83 & 158\\
Scenario 5 & 1.0 & 0.56 - 0.8 & 239\\
Scenario 6 & 1.0 & 0.69 - 1.0 & 187\\
Scenario 7 & 1.0 & 0.56 - 1.0 & 251\\
Scenario 8 & 1.0 & 0.83 - 1.0 & 273\\
\hline
\textbf{Overall} & \textbf{0.94} & \textbf{0.64 - 0.89} & \textbf{223}\\

\hline
\end{tabular}
\caption{End-to-end evaluation.}
\label{tab:e2e_agent_eval_overall}
\vspace{-2mm}
\end{table}
\section{Real-World Network use cases}
\label{sec:swisscom}

This section validates Aether's practical utility on real-world networks from a major ISP. We deployed Aether on a laboratory replica of the ISP's production environment---scaled in device count but preserving all architectural patterns, protocol interactions, and policy complexity. The test network includes $25$ routers across CORE, Aggregation, and Metro layers, with $277$ IPv4/IPv6 addresses, $263$ VRFs across $613$ instances, and $76$ ACL lists containing $274$ rules---totaling over $30,000$ lines of production-equivalent configuration.

We evaluate Aether on two real past incidents caused by faulty change candidates. Unlike synthetic benchmarks, these involve production-type topologies, real vendor configurations (Cisco IOS-XR), and subtle protocol interactions. Ground truth is established by network experts with incident post-mortem input. 

\textbf{Scenario 1: IS-IS Redistribution Loop.} Bidirectional route redistribution between two IS-IS domains used \texttt{metric- type internal}, causing redistributed routes to be re-imported into their originating domain, creating routing loops. The configuration bypassed loop prevention by treating redistributed routes as native IS-IS routes. Ground truth includes end-to-end IPv6 connectivity verification, validation of standard metric configurations to
prevent loops, and confirmation of active IS-IS redistribution.

\textbf{Scenario 2: SRv6 Prefix Summarization Blackhole.} IPv6 prefix summarization was migrated between aggregation routers at L1/L2 boundaries. The same summary prefix was configured on a new router with different backing prefixes. Core routers load-balanced via ECMP, but $50$\% of flows were blackholed at the new aggregation router, which lacked backing routes.
Ground truth includes verification of summarized route advertisements, negative assertions for suppressed specific routes, and regression testing for unrelated traffic flows.
Evaluation metrics are reported in Table~\ref{tab:usecase-metrics}.

\noindent\textbf{Error Detection Accuracy} -- Both scenarios achieve high detection ($100$\%): bad candidates are consistently rejected. Precision ranges from 57\% (Scenario 1) to 90\% (Scenario 2). The lower precision reflects conservative agent behavior, flagging ambiguous behaviors as risks.

 Despite variable precision, 100\% coverage of expert requirements combined with granular multi-test verification ensures robustness: even if one check fails, others covering the same requirement detect the bug. Tool-based verifications significantly outperformed query-based approaches; the majority of precision issues stemmed from the NDM Query Agent either misforging complex AQL queries or misinterpreting results. The SRv6 blackhole case was particularly challenging due to overlapping prefixes and partial summarization.

\noindent\textbf{Test Plan Coverage} -- Aether achieves 100\% cumulative coverage, proving its capability to generate all required tests. In individual executions, consistency remains high with an average coverage of 91.7\% for IS-IS and 95.6\% for SRv6 scenarios. The primary omissions are specific test types: the end-to-end connectivity check is missed in 25\% of IS-IS runs, and the negative routing assertion (verifying routes are not advertised) is missed in 11\% of SRv6 runs. 96-98\% of these tests were relevant to specific requirements, showing the ability of the agent also to avoid wasting resources on useless tests, or more likely to fail in case of hallucinations. Finally, redundancy ranging from 1.7\% to 24.6\%, proved to be mostly due to the generation of multiple granular verification of the same property from different perspective (eg. verify summary prefix  present, and verify specific prefixes not present), allowing more robustness but at the same time being more challenging for precision.

\noindent\textbf{Time to Answer.} The average execution time per run was short (approx. 6 minutes), significantly faster than manual lab testing, and validating the workflow's suitability for CI/CD pipelines. In our experiments, 55\% of the total runtime was dedicated to tool execution—an incompressible cost of network verification. Agentic reasoning accounted for the remaining 45\%: the \textit{Test Planner} (2\%) and \textit{Test Executor} (12\%) remain lightweight, while the most complex tasks are handled by the \textit{Impact Assessment} agent (18\%) and the \textit{NDM Query} agent, which dominates the compute budget (68\%). Despite having room for improvement (eg. query generation; parallel test execution), this confirms that the agentic component remains scalable -- it grows with test plan size instead -- making the approach a viable replacement for slow manual procedures.

\section{Discussion}
\label{sec:discussion}
Our evaluation demonstrates that agentic AI with NDT effectively streamlines network change validation. Aether detects up to 94\% of issues, including intent deviations missed by traditional automation, and achieves 100\% detection of critical protocol-interaction issues in real-world ISP validations.
The key takeaways are summarized below:

\textbf{Intent-Aware Verification.} Standard CI/CD pipelines face an efficiency dilemma: full-matrix checks are prohibitive, while sparse tests miss intent-specific failures. In Scenario 2 (SRv6 migration), generic ECMP tests would pass 50\% of the time despite blackholes. Aether's intent-aware approach generated precise validations---verifying exact summary prefixes and absence of specific routes---demonstrating differential verification impossible without understanding the change's purpose. Agents effectively leveraged model-based tools like Batfish, extending their expressiveness through reasoning and composition.

\textbf{Synergy of Tools and Reasoning.} Effective verification relies on complementary strengths of NDM and NDT tools. The Knowledge Graph provides granular configuration data inaccessible to broad verification tools, yet direct verification proved fragile due to OpenConfig schema complexity. Conversely, NDT tools are essential for verifying complex states like loops or reachability. Future improvements require enhancing agent graph interaction through specialized skills and schema enforcement.

\textbf{Knowledge as an Artifact.} Specialized knowledge injection is critical. Protocol interactions, use-case patterns, and optimization guidelines must be explicit "skills" or memory. Building such a knowledge base is a high-value artifact, essential for modern network complexity that general-purpose models cannot derive from zero-shot reasoning.

\textbf{Operational Value.} The ISP use case evaluation confirms Aether's role as a safety layer. With 100\% bug detection and high expert coverage, it catches critical failures bypassing standard controls. The 73\% global precision indicates acceptable overhead and provides a robust safety net. Future work will focus on multi-agent collaboration and memory-based learning to improve efficiency.

\textbf{Evaluation Rigor.} Ground truth creation remains challenging, requiring iterative expert validation. Aether demonstrates that agents can autonomously execute routine validations, freeing operators for strategic oversight, though quantitative measures for diagnostic accuracy and expanded benchmarks remain needed.

\begin{table}[t]
\centering
\begin{tabularx}{\columnwidth}{@{} X ll @{}}
\toprule
\textbf{Metric} & \textbf{IS-IS Loop} & \textbf{SRv6 Blackhole} \\
\hline
Error Detection & 100\% & 100\%\\
Precision & 57\% & 90\% \\
Main Error Precision & 100\% & 100\%\\
\midrule
Coverage        & 91.7\%               & 95.6\%                   \\
\addlinespace[2pt]
Efficiency      & 98.3\%              & 95.7\%                  \\
\addlinespace[2pt]
Redundancy      & 1.7\%               & 24.6\%                  \\
\midrule
Time to Answer  & 414s & 395s          \\
\bottomrule
\end{tabularx}
\caption{Real-world use cases evaluation}
\label{tab:usecase-metrics}
\end{table}

\section{Conclusions}
\label{sec:concl}

In this paper we present Aether, an agentic system combining five NetOps-specialized generative AI agents with a multi-functional Network Digital Twin for network change validation. We evaluate Aether on $8$ network change scenarios and $2$ operational use cases in a major ISP's network, demonstrating promising results in error detection, coverage, consistency, and robustness compared to traditional CI/CD pipelines.
Aether's effectiveness stems from its neuro-symbolic design: LLM-based agents generate and orchestrate verification workflows while reasoning over ambiguous inputs, paired with structured network representation and specialized verification tools from the NDT. This separation enables flexible intent interpretation with correct and consistent validation execution.

This work makes three primary contributions: (1) a neuro-symbolic architecture bridging natural language intent with formal verification through LLMs, Knowledge Graphs, and verification tools; (2) a comprehensive evaluation methodology assessing individual agent and system performance (Correctness, Consistency, Robustness, Error Detection, Precision), complemented by a thorough assessment of test plan quality in production environments (Coverage, Efficiency, Redundancy); and (3) empirical evidence from 8 synthetic and 2 production scenarios that intent-aware validation detects logic errors missed by regular CI/CD pipelines, and fully leverages a wide range of tooling including model-based verifiers, achieving $94\%$ Error Detection and $64\%$ Precision in synthetic benchmarks, and $100\%$ detection with $73\%$ Precision in production environments.
Aether demonstrates practical value for autonomous network operations, serving as an intelligent assistant for routine validation while operators focus on strategic decisions. Future work will focus on: (1) protocol-specific knowledge injection via skills and memory-based learning; (2) multi-agent collaboration with autonomous agent workflows; (3) feedback loops from production failures; and (4) operational tooling for false positive management.
\newpage
\bibliographystyle{IEEEtran}
\bibliography{arxivreference}

@inproceedings{ns3,
  author    = {Riley, George F. and Henderson, Thomas R.},
  title     = {The ns-3 Network Simulator},
  booktitle = {Modeling and Tools for Network Simulation},
  year      = {2010},
  publisher = {Springer},
  address   = {Berlin, Heidelberg},
  pages     = {15--34},
  doi       = {10.1007/978-3-642-12331-3_2}
}

@software{Containerlab,
  author  = {Sukharev, Roman and Contributors},
  title   = {Containerlab},
  url     = {https://github.com/srl-labs/containerlab},
  year    = {2025},
  version = {v0.x},
}

@misc{agntcy,
  author       = {{Linux Foundation}},
  title        = {AGNTCY Project: Building infrastructure for the Internet of Agents. },
  howpublished = {\url{https://agntcy.org/}},
  year         = {2025},
  month        = jul
}

@misc{AAIF,
  author       = {{Linux Foundation}},
  title        = {Agentic AI Foundation (AAIF): Advancing Agentic AI Together.},
  howpublished = {\url{https://aaif.io/}},
  year         = {2025},
  month        = dec,
  note         = {AAIF establishes neutral, open governance for agentic AI standards including MCP, goose, and AGENTS.md}
}

@techreport{TMForum,
  author       = {{TM Forum}},
  title        = {Digital Twin for Decision Intelligence (DT4DI): From Strategy to Implementation},
  institution  = {TM Forum},
  year         = {2025},
  number       = {IG1307},
  url          = {https://www.tmforum.org/resources/introductory-guide-whitepaper/dt4di-from-strategy-to-implementation-v3-0-0-ig1307/},
}

@article{ndt_survey,
  author  = {Raza, Syed Mohsan and Minerva, Roberto and Crespi, Noel and Alvi, Maira and Herath, Manoj and Dutta, Hrishikesh},
  title   = {A Comprehensive Survey of Network Digital Twin Architecture, Capabilities, Challenges, and Requirements for Edge-Cloud Continuum},
  journal = {Computer Communications},
  year    = {2025},
  publisher = {Elsevier},
  url     = {https://hal.science/hal-04986834},
}

@misc{GNS3,
  author       = {{GNS3 Development Team}},
  title        = {GNS3 Network Emulator},
  howpublished = {\url{https://www.gns3.com}},
  year         = {2025},
  note         = {Graphical network simulator supporting real and virtual devices},
}

@article{omnet,
  author  = {Varga, Andr{\'a}s and Hornig, Rudolf},
  title   = {An Overview of the OMNeT++ Simulation Environment},
  journal = {Proceedings of the 1st International Conference on Simulation Tools and Techniques (Simutools)},
  year    = {2008},
  pages   = {1--10},
  publisher = {ACM},
  doi     = {10.4108/ICST.SIMUTOOLS2008.3027}
}

@inproceedings{hotnets25,
  author    = {Krentsel, Alexander and Ye, Oliver and Tafoya, Anthony and Ma, Xuqian and Ratnasamy, Sylvia and Shaikh, Anees},
  title     = {Towards Accessible Model-Free Verification},
  booktitle = {Proceedings of the 24th ACM Workshop on Hot Topics in Networks (HotNets '25)},
  year      = {2025},
  month     = nov,
  address   = {College Park, MD, USA},
  publisher = {ACM},
  doi       = {10.1145/3772356.3772380},
  url       = {https://conferences.sigcomm.org/hotnets/2025/papers/hotnets25-final13.pdf},
}

@misc{openconfig,
  author       = {OpenConfig Working Group},
  title        = {OpenConfig Data Model},
  howpublished = {\url{https://www.openconfig.net/}},
  note         = {Accessed: 2025-10-06}
}

@inproceedings{Datalog,
  author = {Li, Yahui and Wang, Zhiliang and Yin, Xia et al.},
  year   = {2018},
  month  = {04},
  pages  = {1-2},
  title  = {Efficient network configuration verification using optimized datalog},
  doi    = {10.1109/INFCOMW.2018.8406876}
}

@inproceedings{HSA,
  author    = {Peyman Kazemian and George Varghese and Nick McKeown},
  title     = {Header Space Analysis: Static Checking for Networks},
  booktitle = {9th USENIX Symposium on Networked Systems Design and Implementation (NSDI 12)},
  year      = {2012},
  isbn      = {978-931971-92-8},
  address   = {San Jose, CA},
  pages     = {113--126},
  url       = {https://www.usenix.org/conference/nsdi12/technical-sessions/presentation/kazemian},
  publisher = {USENIX Association},
  month     = apr
}

@inproceedings{Azure1,
  author    = {Bj{\o}rner, Nikolaj
               and Jayaraman, Karthick},
  editor    = {Natarajan, Raja
               and Barua, Gautam
               and Patra, Manas Ranjan},
  title     = {Checking Cloud Contracts in Microsoft Azure},
  booktitle = {Distributed Computing and Internet Technology},
  year      = {2015},
  publisher = {Springer International Publishing},
  address   = {Cham},
  pages     = {21--32},
  isbn      = {978-3-319-14977-6}
}

@inproceedings{Azure2,
  author    = {Beckett, Ryan and Gupta, Aarti and Mahajan, Ratul and Walker, David},
  title     = {A General Approach to Network Configuration Verification},
  booktitle = {Proc. of SIGCOMM '17, Los Angeles, CA, USA, August 21-25, 2017},
  year      = {2017},
  month     = {August},
  publisher = {ACM - Association for Computing Machinery},
  pages     = {14},
  edition   = {Proc. of SIGCOMM ’17, Los Angeles, CA, USA, August 21-25, 2017}
}

@inproceedings{Azure3,
  author       = {Beckett, Ryan and Mahajan, Ratul},
  title        = {Putting network verification to good use},
booktitle = {Proc. of ACM HotNets '19, Princeton, NJ, USA, November 13-15, 2019},
  organization = {ACM},
  month        = {November},
}

@inproceedings{namrud2024kubeplaybook,
  title={Kubeplaybook: A repository of ansible playbooks for kubernetes auto-remediation with llms},
  author={Namrud, Zakeya and Sarda, Komal and Litoiu, Marin et al.},
  booktitle={Companion of the 15th ACM/SPEC International Conference on Performance Engineering},
  pages={57--61},
  year={2024}
}

@inproceedings{mondal2023llms,
  title={What do llms need to synthesize correct router configurations?},
  author={Mondal, Rajdeep and Tang, Alan and Beckett, Ryan and Millstein, Todd and Varghese, George},
  booktitle={Proc. of ACM HotNets, Cambridge, MA, USA, November 28-29, 2023},
  pages={189--195},
  year={2023}
}

@misc{react,
      title={ReAct: Synergizing Reasoning and Acting in Language Models}, 
      author={Shunyu Yao and Jeffrey Zhao and Dian Yu and Nan Du and Izhak Shafran and Karthik Narasimhan and Yuan Cao},
      year={2023},
      eprint={2210.03629},
      archivePrefix={arXiv},
      primaryClass={cs.CL},
      url={https://arxiv.org/abs/2210.03629}, 
}

@inproceedings{sahoo2024ansible,
  title={Ansible lightspeed: A code generation service for it automation},
  author={Sahoo, Priyam and Pujar, Saurabh and Nalawade, Ganesh and Genhardt, Richard and Mandel, Louis and Buratti, Luca},
  booktitle={Proc. of IEEE/ACM International Conference on Automated Software Engineering},
  pages={2148--2158},
  year={2024}
}

@misc{simap,
  title  = {IETF NMOG Internet Draft - SIMAP: Concept, Requirements, and Use Cases},
  author = {Havel, O. and Claise, B. and Dios, O.G.D. and Graf, T.},
  month  = {Oct},
  year   = {2025},
  howpublished = {\url{https://datatracker.ietf.org/doc/draft-ietf-nmop-simap-concept/}},
}

@misc{twineu,
  title  = {Developing a concept of Pan-European Digital Twin of the electricity system},
  author = {{Twin EU Consortium}},
  month  = {Jan},
  year   = {2024},
  howpublished = {\url{https://twineu.net/}},
}

@misc{SLIM,
    title = {IETF draft-mpsb-agntcy-slim-00, Secure Low-latency Interactive Messaging},
    author = {{IETF AGNTCY Working Group}},
    year = {2024},
    howpublished = {\url{https://datatracker.ietf.org/doc/draft-mpsb-agntcy-slim/}},
}

@misc{Oxford-Economics,
  title = {The hidden costs of downtime:The \$400B problem facing the Global 2000},
  author = {Mohanty, A. and Robinson, T. and O'Farrell, A.},
  month  = {Jul},
  year   = {2024},
  howpublished = {\url{https://www.oxfordeconomics.com/resource/the-hidden-costs-of-downtime-the-400b-problem-facing-the-global-2000/}},
}

@inproceedings{Alibaba1,
  author    = {Ye, Fangdan and Yu, Da and Zhai, Ennan and Liu, Hongqiang Harry et al.},
  title     = {Accuracy, Scalability, Coverage: A Practical Configuration Verifier on a Global WAN},
  year      = {2020},
  isbn      = {9781450379557},
  publisher = {ACM},
  address   = {New York, NY, USA},
  doi       = {10.1145/3387514.3406217},
  booktitle = {Proc. of ACM SIGCOMM 2020},
  pages     = {599–614},
  numpages  = {16},
  location  = {Virtual Event, USA},
  series    = {SIGCOMM '20}
}

@inproceedings{Alibaba3,
  author    = {Zheng, Naiqian and Liu, Mengqi and Zhai, Ennan et al.},
  title     = {Meissa: scalable network testing for programmable data planes},
  year      = {2022},
  isbn      = {9781450394208},
  publisher = {Association for Computing Machinery},
  address   = {New York, NY, USA},
  booktitle = {Proc. of ACM SIGCOMM 2022},
  pages     = {350–364},
  numpages  = {15},
  location  = {Amsterdam, Netherlands},
  series    = {SIGCOMM '22}
}

@inproceedings{Alibaba4,
  author    = {Tian, Bingchuan and Gao, Jiaqi and Liu, Mengqi et al.},
  title     = {Aquila: a practically usable verification system for production-scale programmable data planes},
  year      = {2021},
  isbn      = {9781450383837},
  publisher = {ACM},
  address   = {New York, NY, USA},
  booktitle = {Proc. of ACM SIGCOMM 2021 },
  pages     = {17–32},
  numpages  = {16},
  location  = {Virtual Event, USA},
  series    = {SIGCOMM '21}
}

@article{Alibaba2,
  author     = {Wang, Kun and Zhao, Chengcheng and Chu, Jinpei and Shi, Yiping and Lu, Jianyuan and Lyu, Biao and Zhu, Shunmin and Cheng, Peng and Chen, Jiming},
  title      = {LFVeri: Network Configuration Verification for Virtual Private Cloud Networks},
  year       = {2024},
  issue_date = {Dec. 2024},
  publisher  = {IEEE Press},
  volume     = {32},
  number     = {6},
  issn       = {1063-6692},
  url        = {https://doi.org/10.1109/TNET.2024.3469386},
  doi        = {10.1109/TNET.2024.3469386},
  journal    = {IEEE/ACM Trans. Netw.},
  month      = oct,
  pages      = {5475–5490},
  numpages   = {16}
}

@inproceedings{rela,
  author    = {Xu, Xieyang and Yuan, Yifei and Kincaid, Zachary and Krishnamurthy, Arvind and Mahajan, Ratul and Walker, David and Zhai, Ennan},
  title     = {Relational Network Verification},
  year      = {2024},
  isbn      = {9798400706141},
  publisher = {Association for Computing Machinery},
  address   = {New York, NY, USA},
  url       = {https://doi.org/10.1145/3651890.3672238},
  doi       = {10.1145/3651890.3672238},
  booktitle = {Proceedings of the ACM SIGCOMM 2024 Conference},
  pages     = {213–227},
  numpages  = {15},
  location  = {Sydney, NSW, Australia},
  series    = {ACM SIGCOMM '24}
}

@inproceedings{brown23lessons,
  author    = {Brown, Matt and Fogel, Ari and Halperin, Daniel and Heorhiadi, Victor and Mahajan, Ratul and Millstein, Todd},
  title     = {Lessons from the evolution of the Batfish configuration analysis tool},
  year      = {2023},
  isbn      = {9798400702365},
  publisher = {Association for Computing Machinery},
  address   = {New York, NY, USA},
  booktitle = {Proceedings of the ACM SIGCOMM 2023 Conference},
  pages     = {122--135},
  numpages  = {14},
  location  = {New York, NY, USA},
  series    = {ACM SIGCOMM '23}
}

@misc{enns2011rfc,
  title={RFC 6241: Network configuration protocol (NETCONF)},
  author={Enns, Rob and Bjorklund, M and Schoenwaelder, J and Bierman, A},
  year={2011},
  publisher={RFC Editor}
}

@misc{ansible,
  author       = {{Red Hat, Inc.}},
  title        = {Ansible: Simple, Agentless IT Automation},
  howpublished = {\url{https://www.ansible.com/}},
  year         = {2024},
  note         = {Open-source automation tool for configuration management and application deployment}
}

@misc{puppet,
  author       = {{Puppet, Inc.}},
  title        = {Puppet: Infrastructure Automation for Security and Compliance},
  howpublished = {\url{https://puppet.com/}},
  year         = {2024},
  note         = {Configuration management and automation platform}
}

@inproceedings{mahajan2002bgw,
  author    = {Mahajan, Ratul and Wetherall, David and Anderson, Tom},
  title     = {Understanding BGP misconfiguration},
  booktitle = {Proceedings of the 2002 Conference on Applications, Technologies, Architectures, and Protocols for Computer Communications (SIGCOMM '02)},
  year      = {2002},
  pages     = {3--16},
  publisher = {ACM},
  address   = {New York, NY, USA},
  doi       = {10.1145/633025.633027}
}

@inproceedings{oppenheimer2003internet,
  author    = {Oppenheimer, David and Ganapathi, Archana and Patterson, David A.},
  title     = {Why do Internet services fail, and what can be done about it?},
  booktitle = {Proceedings of the 4th USENIX Symposium on Internet Technologies and Systems (USITS '03)},
  year      = {2003},
  pages     = {1--16},
  publisher = {USENIX Association}
}

@inproceedings{rusek2020routenet,
  author    = {Rusek, Krzysztof and Su{\'a}rez-Varela, Jos{\'e} and Mestres, Albert and Barlet-Ros, Pere and Cabellos-Aparicio, Albert},
  title     = {RouteNet: Leveraging Graph Neural Networks for network modeling and optimization in SDN},
  booktitle = {IEEE Journal on Selected Areas in Communications},
  year      = {2020},
  volume    = {38},
  number    = {10},
  pages     = {2260--2270},
  publisher = {IEEE},
  doi       = {10.1109/JSAC.2020.3000405}
}

@inproceedings{netcov_sdn_coverage,
  author    = {Kazemian, Peyman and Chang, Michael and Zeng, Hongyi and Varghese, George and McKeown, Nick},
  title     = {Real Time Network Policy Checking Using Header Space Analysis},
  booktitle = {Proceedings of the 10th USENIX Symposium on Networked Systems Design and Implementation (NSDI)},
  year      = {2013},
  address   = {Lombard, IL, USA}
}

@misc{becket2020nv2,
  author       = {Beckett, Ryan and Mahajan, Ratul},
  title        = {Network verification 2.0},
  howpublished = {\url{https://netverify.fun/network-verification-2-0/}}, 
  year         = {2020},
  month        = {dec}
}

@misc{a2a_spec_2025,
  author = {{Linux Foundation}},
  title = {Agent-to-Agent (A2A) Protocol Specification v1.0},
  year = {2025},
  url = {https://a2a-protocol.org/latest/specification/},
  note = {Accessed: 2026-02-02}
}

@misc{liu2023llamaindex,
  author = {Jerry Liu and LlamaIndex Contributors},
  title = {LlamaIndex: A Data Framework for LLM Applications},
  year = {2023},
  howpublished = {\url{https://github.com/run-llama/llama_index}},
}

@misc{mcp_spec_2025,
  author = {{Model Context Protocol Community}},
  title = {Model Context Protocol (MCP) Specification},
  year = {2025},
  url = {https://modelcontextprotocol.io/specification/},
  note = {Accessed: 2026-02-02}
}

@article{NetKAT,
author = {Xu, Han and Kincaid, Zachary and Mahajan, Ratul and Walker, David},
title = {Network Change Validation with Relational NetKAT},
year = {2026},
issue_date = {January 2026},
publisher = {ACM},
address = {New York, NY, USA},
volume = {10},
number = {POPL},
journal = {Proc. ACM Program. Lang.},
month = jan,
articleno = {14},
numpages = {29}
}

@software{deepeval,
  author       = {{Confident AI}},
  title        = {DeepEval: The LLM Evaluation Framework},
  year         = {2026},
  url          = {https://github.com/confident-ai/deepeval},
  note         = {Version 3.8.1. Open-source evaluation framework for LLMs.},
}

@inproceedings{batfish,
  author    = {Guha, Arjun and Reitblatt, Mark and Foster, Nate},
  title     = {Batfish: Analyzing Network Configurations Using a Data Plane Abstraction},
  booktitle = {Proceedings of the 12th USENIX Symposium on Networked Systems Design and Implementation (NSDI)},
  year      = {2015},
  pages     = {283--296},
  publisher = {USENIX Association},
}

@misc{arangodb,
  author       = {{ArangoDB GmbH}},
  title        = {ArangoDB: The Native Multi-Model Database},
  howpublished = {\url{https://www.arangodb.com}},
  year         = {2025},
  note         = {Open-source native multi-model database supporting graph, document, and key--value data models}
}

@article{kakarla2024diffy,
  title={Diffy: Data-Driven Bug Finding for Configurations},
  author={Kakarla, Siva Kesava Reddy and Yan, Francis Y and Beckett, Ryan},
  journal={Proceedings of the ACM on Programming Languages},
  volume={8},
  number={PLDI},
  pages={199--222},
  year={2024},
  publisher={ACM New York, NY, USA}
}

@article{xu2026network,
  title={Network Change Validation with Relational NetKAT},
  author={Xu, Han and Kincaid, Zachary and Mahajan, Ratul and Walker, David},
  journal={Proceedings of the ACM on Programming Languages},
  volume={10},
  number={POPL},
  pages={384--412},
  year={2026},
  publisher={ACM New York, NY, USA}
}

@manual{cisco_nso_2024,
  title        = {Cisco Network Services Orchestrator (NSO) Documentation},
  author       = {{Cisco Systems, Inc.}},
  organization = {Cisco Systems, Inc.},
  year         = {2024},
  note         = {Version 6.x},
  url          = {https://developer.cisco.com/docs/nso/},
}

\end{document}